\documentclass{aastex63}

\usepackage{tikz}
\usetikzlibrary{arrows}
\usepackage{amsmath}

\submitjournal{APJ}
\accepted{\today}

\shorttitle{Diffuse Reflection with Thermal Emission}
\shortauthors{Sengupta et al.}

\graphicspath{{./}{figures/}}

\begin{document}

\title{Effects of thermal emission on Chandrasekhar's semi-infinite diffuse reflection problem}

\correspondingauthor{Soumya Sengupta}
\email{soumya.s@iiap.res.in}

\author[0000-0002-7006-9439]{Soumya Sengupta}
\affiliation{Indian Institute of Astrophysics, Koramangala 2nd Block,
Sarjapura Road, Bangalore 560034, India}
\affiliation{Pondicherry University, R.V. Nagar, Kalapet, 605014, Puducherry, India}


\section{Abstract}
\textbf{Context}: The analytical results of Chandrasekhar's semi-infinite diffuse reflection problem is crucial in the context of stellar or planetary atmosphere. However, the atmospheric emission effect was not taken into account in this model, and the solutions are applicable only for diffusely scattering atmosphere in absence of emission.

\textbf{Aim}:  We extend the model of semi-infinite diffuse reflection problem by including the effects of \textit{thermal emission} $B(T)$, and present how this affects Chandrasekhar's analytical end results. Hence, we aim to generalize Chandradekhar’s model to provide a complete picture of this problem.

\textbf{Method}: We use \textit{Invariance Principle Method} to find the radiative transfer equation accurate for diffuse reflection in presence of $B(T)$. Then we derive the modified scattering function $S(\mu,\phi;\mu_0,\phi_0)$ for different kind of phase functions . 

\textbf{Results}:  We find that, the scattering function $S(\mu,\phi;\mu_0,\phi_0)$ as well as diffusely reflected specific intensity $I(0,\mu;\mu_0)$ for different phase functions are modified due to the emission $B(T)$ from layer $\tau=0$. In both cases, $B(T)$ is added to the results of only scattering case derived by Chandrasekhar, with some multiplicative factors. Thus the diffusely reflected spectra will be enriched and carries the temperature information of $\tau=0$ layer. As the effects are additive in nature, hence our model reduces to the sub-case of Chandrasekhar's scattering model in case of $B(T)=0$. We conclude that our generalized model provides more accurate results due to the inclusion of the thermal emission effect in Chandrasekhar's semi-infinite atmosphere problem.


\section{Introduction}\label{intro}

The study of analytical radiative transfer theory is a very important and well established branch of physics. In mid 20th century, while developing the analytical formulation of this theory, Chandrasekhar introduced the well known semi-infinite atmosphere problem. Almost at the same time Ambartsumian introduced his well known invariance principle \citet{ambartsumian1943cr} and \citet{ambartsumian1944problem}.
For diffuse reflection case the principle is reformulated by \citet{Chandrasekhar} as, \textit{the law of diffuse reflection by a homogeneous semi-infinite plane-parallel atmosphere must be invariant to the addition (or subtraction) of layers of arbitrary optical thickness to (or form) the atmosphere}.\citet{chandrasekhar1947radiative} analytically solved the semi-infinite atmosphere problem and it reduced in terms of the well known Chandrasekhar's Scattering function (S) and H-function. The same treatment has been done for different scattering phase functions by \citet{horak1950diffuse},\citet{horak1961diffuse},\citet{abhyankar1970imperfect}. For a full review on analytical radiative transfer problem, see \citet{rybicki1996radiative}.

The analytic solutions of radiative transfer given by  \citet{Chandrasekhar} has a large range of direct applications, starting from planetary atmosphere modelling \citet{madhusudhan2012analytic}, to ion induced secondary electron emission \citet{dubus1986theoretical}. However, the solutions of semi-infinite atmophere problem \citet{chandrasekhar1947radiative} is obtained specifically for a diffusely reflcting atmosphere without atmospheric emission. Although the emission effect has been studied for planetary atmosphere problem \citet{bellman1967chandrasekhar}, \citet{grant1968solution},\citet{domanus1974fundamental} the semi-infinite atmosphere problem with emission effect remains unsolved.

In plane-parallel radiative transfer equation, the total radiation added by each atmospheric layer to the transfer equation is known as the \textit{Source function}. The diffused reflection as well as emission from each layer both affect the Source function  \citet{domanus1974fundamental}. Recently, using diffused reflection and transmission model, \citet{sengupta2020optical} demonstrated the crucial effect of scattering on transmitted flux for hot-jupiter atmospheres, while 
\citet{chakrabarty2020effects} showed the significant effect of thermal re-emission process in upper atmosphere for the same. Thus for a complete solution of semi-infinite diffuse reflection problem, the inclusion of scattering as well as emission is important.

To include the emission effect in semi-infinite atmosphere problem we consider the simplest case of homogeneous atmosphere whose layers are in \textit{Local Thermodynamic Equilibrium}. In this case each atmospheric layer emits a planck emission $B(T_\tau)$ depending only on the layer temperature $T_\tau$  \citet{seager2010exoplanet}. Thus the model not only reveals the scattering properties of the layer but also carries the temperature information. Again the resultant radiation $I(0,\mu;\mu_0)$ is increased by the inclusion of thermal emission over the scattering only case. Moreover the model presented here is more generalized and accurate than the scattering only reflection model.

 In section~\ref{transfer equation} we present the derivation of radiative transfer equation appropriate for diffuse reflection in presence of thermal emission using invariance principle method. We derive the integral equation of scattering function in case of thermally emitting semi-infinite atmosphere in section~\ref{integral form of scattering function}. Section~\ref{S with different p} will show the explicit form of the integral equation for different cases of \textit{Isotropic scattering, Asymmetric scattering, Rayleigh scattering} and the general scattering phase function with $p(\cos\Theta)=\tilde{\omega_0}+\tilde{\omega_1}P_1(\cos\Theta)+\tilde{\omega_2}P_2(\cos\Theta)$. The comparison of end results in semi-infinite atmosphere problem in presence of thermal emission (our work) and in absence of it \citet{Chandrasekhar} is given in section~\ref{comparison with chandrasekhar}. Section~\ref{B(T) contribution} is devoted on explaining how thermal emission actually effects the analytic end results. Finally, we interpret our results and suggest future work in the last section.

	
\section{Derivation of Diffusion Transfer equation in presence of Thermal Emission}\label{transfer equation}
	The radiative transfer equation in case of plane parallel approximation can be written as,

\begin{equation}\label{plane parallel radiative transfer equation}
\mu\frac{dI(\tau,\nu,\mu,\phi)}{d\tau_\nu} =I(\tau,\nu,\mu,\phi)-\xi(\tau,\nu,\mu,\phi)
\end{equation}

Here, $I(\tau,\nu,\mu,\phi)$ is the specific intensity at frequency $\nu$ along $\mu$ , the direction cosine of inclination angle with the outward normal of atmospheric layer and $\phi$ is the azimuthal angle. Here, $d\tau_\nu$ is the \textit{optical thickness} defined as \citet{seager2010exoplanet},
\begin{equation}\label{optical depth}
d\tau_\nu(z) = -\kappa_\nu(z) dz
\end{equation}

where, z is the atmospheric height. $\kappa(\tau,\nu)$ and $\xi(\tau,\nu)$ are the volumetric absorption co-efficient and source function respectively at the frequency $\nu$ along the direction $\mu$, from the atmospheric layer with optical depth $\tau$. A
semi-infinite atmosphere is bounded on one side at $\tau=0$ and extends upto $\infty$ on the other direction \citet{Chandrasekhar}.

Chandrasekhar derived analytic solutions for the semi-infinite diffuse reflection problem of a \textit{scattering only  atmosphere} \citet{chandrasekhar1947radiative}. We will introduce the atmospheric emission along with the scattering. Thus for an atmosphere with both emission as well as scattering,  the atmospheric extinction can be characterized in terms of volumetric \textit{absorption co-efficient $\kappa$, scattering co-efficient $\sigma$} and \textit{extinction co-efficient $\chi$} which follows the relation given by, \citet{domanus1974fundamental} and \citet{sengupta2020optical} 
\begin{equation}\label{chi relation}
\chi(\tau,\nu) = \kappa(\tau,\nu) + \sigma(\tau,\nu)
\end{equation}

So, eqn.\eqref{optical depth} will change as,
\begin{equation}\label{optical depth1}
d\tau_\nu = -\chi(z,\nu) dz
\end{equation}

We mention that the radiation field $I(\tau,\nu,\mu,\phi)$ at optical depth $\tau$ is the sum of the diffuse radiation $I_D(\tau,\nu,\mu,\phi)$ and an attenuated field along the direction of incident radiation. It can be expressed mathematically using  Dirac-delta function ($\delta$) as,
\begin{equation}
\begin{split}
I(\tau,\nu,+\mu,\phi) = I_D(\tau,\nu,+\mu,\phi) \hspace{2.5cm} (0<\mu\leqslant1)\\
I(\tau,\nu,-\mu,\phi) = I_D(\tau,\nu,-\mu,\phi) + \pi F \delta(\mu-\mu_0)\delta(\phi-\phi_0) \hspace{1cm} (0<\mu\leqslant1)
\end{split}
\end{equation}
following \cite{horak1961diffuse} and the boundary conditions therein. We will drop the subscript D for the remaining paper and I denotes the intensity for diffused radiation only.


\begin{figure}[h!]\label{drawing source function}
\begin{center}
\begin{tikzpicture}
\draw (0,0)--(9,0)node[right] {$\tau=0$};
\draw (0,-2)--(9,-2)node[right]{$\tau=\tau_1$};
\draw[->] (3.9,1.9)--(3.4,0.85);
\draw (3.4,1.6)--(4.0,1.4);
\draw (3.4,0.85)--(3,0);
\draw[<-] (3.7,0.2)--(3,0);
\draw[<-] (3.5,0.5)--(3,0);
\draw[<-] (3,0.6)--(3,0);
\draw[<-] (2.5,0.5)--(3,0);
\draw[<-] (2.3,0.2)--(3,0);
\draw[<-] (6.7,0.2)--(6,0);
\draw[<-] (6.5,0.5)--(6,0);
\draw[<-] (6,0.6)--(6,0);
\draw[<-] (5.5,0.5)--(6,0);
\draw[<-] (5.3,0.2)--(6,0);
\draw[<-] (6.7,-0.2)--(6,0);
\draw[<-] (6.5,-0.5)--(6,0);
\draw[<-] (6,-0.6)--(6,0);
\draw[<-] (5.5,-0.5)--(6,0);
\draw[<-] (5.3,-0.2)--(6,0);
\node[draw] at (7,1.3) {B($\tau=0$)};
\draw (3,0) -- (3,-2);
\node[draw] at (2,-1){$Fe^{-\frac{\tau}{\mu_0}}$};
\draw[->] (3,-2)--(3.7,-2.2);
\draw[->] (3,-2)--(3.5,-2.5);
\draw[->] (3,-2)--(3,-2.6);
\draw[->] (3,-2)--(2.5,-2.5);
\draw[->] (3,-2)--(2.3,-2.2);
\draw[->] (6,-2)--(6.7,-1.8);
\draw[->] (6,-2)--(6.5,-1.5);
\draw[->] (6,-2)--(6,-1.4);
\draw[->] (6,-2)--(5.5,-1.5);
\draw[->] (6,-2)--(5.3,-1.8);
\draw[->] (6,-2)--(6.7,-2.2);
\draw[->] (6,-2)--(6.5,-2.5);
\draw[->] (6,-2)--(6,-2.6);
\draw[->] (6,-2)--(5.5,-2.5);
\draw[->] (6,-2)--(5.3,-2.2);
\node[draw] at (6.1,-3) {B($\tau=\tau_1$)};
\end{tikzpicture}
\end{center}
\caption{This figure shows the total effect due to diffuse scattering as well as thermal emission for a \textit{semi-infinite atmosphere}. The thermal emission \textit{B} is Isotropic in nature and emitted from the particular atmospheric layer where the optical depth is $\tau$}
\end{figure}
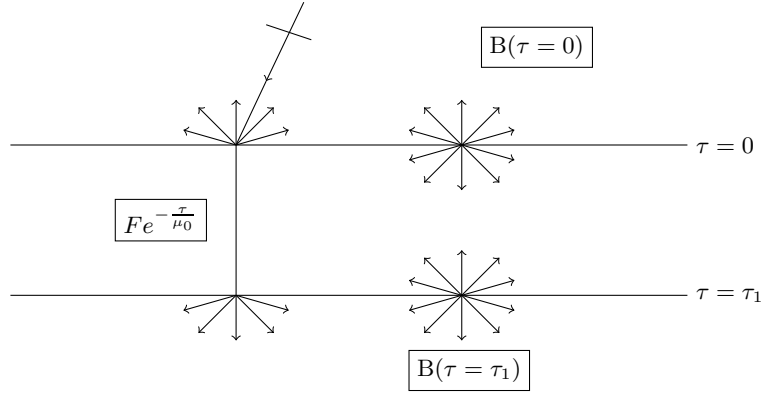

 Here $\pi F$ is the amount of flux incident on a plane-parallel atmospheric layer along the direction ($-\mu_0,\phi_0$), and $\pi F e^{-\frac{\tau}{\mu_0}}$ is the \textit{reduced incident radiation} that penetrates upto the optical depth $\tau$ without experiencing any scattering \citet{Chandrasekhar}.
Also, the diffused radiation scattered at optical depth $\tau$ can be expressed in terms of the incident radiation and phase function $p(\mu,\phi;\mu',\phi')$, which is the angular distribution of photons before (from direction $\mu',\phi'$) and after (to direction $\mu,\phi$) scattering.

Now the contribution of atmospheric emission can be considered in terms of $\beta(\tau,\nu,\mu,\phi)$, which represents the angular distribution of the energy emitted from the atmospheric layer at optical depth $\tau$ and at frequency $\boldsymbol{\nu}$. While considering all of these, the source function can be written as,
\begin{equation}\label{source function for emission+scattering1}
\begin{split}
\xi(\tau,\mu,\phi) =& \beta(\tau,\mu,\phi) +\frac{1}{4\pi}\int_{-1}^1\int_{0}^{2\pi}p(\mu,\phi;\mu',\phi')I(\tau,\mu',\phi')d\phi'd\mu'\\
&+\frac{1}{4}Fe^{-\tau/\mu_0}p(\mu,\phi;-\mu_0,\phi_0)
\end{split}
\end{equation}
From this equtaion onwards we will consider all expressions at a single frequency unless otherwise mentioned and thus drop $\nu$.

The emission, $\beta(\tau,\mu,\phi)$ can be caused by internal energy source distribution \citet{bellman1967chandrasekhar}, the thermal re-emission process \citet{chakrabarty2020effects} or the temperature dependent planck emission \citet{malkevich1963angular}.
However, for an atmosphere satisfying homogeneity and Local Thermodynamic Equilibrium conditions, $\beta(\tau,\mu,\phi)$ can be written in terms of planck function \cite{seager2010exoplanet} as, 
$\beta(\tau,\mu,\phi) = \frac{\kappa}{\chi}B(T_\tau)$. Here $T_\tau$ is the temperature of an atmospheric layer with optical depth $\tau$. Now in the low scattering limit $\kappa>>\sigma$, eqn.\eqref{chi relation} will be reduced into $\chi\approx\kappa$. In this limit, the atmospheric emission $\beta$ will be equivalent to the planck function B. So eqn.\eqref{source function for emission+scattering1} will then have the following form,

\begin{equation}\label{source function for thermal emission+scattering1}
\begin{split}
\xi(\tau,\mu,\phi) =& B(T_\tau) +\frac{1}{4\pi}\int_{-1}^1\int_{0}^{2\pi}p(\mu,\phi;\mu',\phi')I(\tau,\mu',\phi')d\phi'd\mu'\\
&+\frac{1}{4}Fe^{-\tau/\mu_0}p(\mu,\phi;-\mu_0,\phi_0)
\end{split}
\end{equation}

So the radiative transfer equation for \textit{an atmosphere with both emission as well as diffuse reflection} can be written as,
\begin{equation}\label{emission with diffusion approximation}
\begin{split}
\mu\frac{dI(\tau,\mu,\phi)}{d\tau} =& I(\tau,\mu,\phi) -B(T_\tau)
-\frac{1}{4\pi}\int_{-1}^1\int_{0}^{2\pi}p(\mu,\phi;\mu',\phi')I(\tau,\mu',\phi')d\mu'd\phi'\\
&-\frac{1}{4}Fe^{-\tau/\mu_0}p(\mu,\phi;-\mu_0,\phi_0)
\end{split}
\end{equation}

This equation represents \textit{the general form of radiative transfer appropriate for diffuse reflection and transmission, where both scattering as well as atmospheric emission takes place}.

\section{General Integral equation of scattering Function}\label{integral form of scattering function}
The Scattering function $S(\mu,\phi;\mu_0,\phi_0)$ for semi-infinite atmosphere was introduced by \citet{Chandrasekhar}. When $\pi F$ amount of flux incident along the direction $(-\mu_0,\phi_0)$ on a plane-parallel semi-infinite atmosphere, then the diffusely reflected intensity $I(0,+\mu,\phi)$ along $(+\mu,\phi)$ direction can be expressed in terms of scattering function as,
  
\begin{equation}\label{I0_1}
I(0,+\mu,\phi)=\frac{F}{4\mu}S(\mu,\phi;\mu_0,\phi_0)
\end{equation}

It is apparent from eqn.\eqref{I0_1} that the diffused intensity is directly proportional to the incident flux F. \cite{Chandrasekhar} showed that, this is indeed the case while considering \textit{only the atmospheric scattering and no emission}. Thus for an atmosphere without emission, the diffused intensity vanishes when there is no flux incident on the surface (i.e. F=0). But this is not the case for a thermally emitting atmosphere. We will show later in this section (eqn.\eqref{scattering function1}) as well as in section~\ref{B(T) contribution} that the scattering function S for a thermally emitting atmosphere always contains an additive term in the form of $\frac{B(T)}{F}$. Thus, eqn\eqref{I0_1} gives a non-vanishing intensity of radiation even in the absence of incident flux (i.e. F=0) due to the thermal emission of the atmosphere. Therefore we mention that although eqn.\eqref{I0_1} shows apparent proportionality of the emitting radiation to the incident flux, in case of thermally emitting atmosphere the scattering function itself takes care of the fact that the thermal emission remains independent of the incident flux.

 We calculate the general functional form of $S(\mu,\phi;\mu_0,\phi_0)$ in case of diffuse scattering and thermal emission process using \textit{invariance principle method} (see section~\ref{intro}) as formulated in \citet{Chandrasekhar} and applied in \citet{chandrasekhar1947radiative} and \citet{horak1961diffuse}. We use this principle to calculate the outward intensity $I(\tau,+\mu,\phi)$ as follows,
\begin{equation}\label{invariance principle}
I(\tau,+\mu,\phi)=\frac{F}{4\mu}e^{-\frac{\tau}{\mu_0}}S(\mu,\phi;\mu_0,\phi_0) + \frac{1}{4\pi\mu}\int_0^1\int_0^{4\pi}I(\tau,-\mu',\phi') S(\mu,\phi;\mu',\phi') d\mu'd\phi'
\end{equation}

The outward ($0\leqslant\mu\leqslant1$) and inward ($-1\leqslant\mu<0$) intensities are denoted by $I(\tau,+\mu,\phi)$ and $I(\tau,-\mu,\phi)$ respectively with the boundary condition,
\begin{equation}\label{boundary condition}
I(0,-\mu,\phi)=0\hspace{1cm}(0<\mu\leqslant 1)
\end{equation}

 Differentiating eqn.\eqref{invariance principle} at $\tau=0$ and comparing with the transfer equation.\eqref{plane parallel radiative transfer equation} alongwith the boundary condition we will get the scattering function in terms of source function,
 \begin{equation}\label{Scattering function integral1}
\frac{F}{4}(\frac{1}{\mu} + \frac{1}{\mu_0})S(\mu,\phi;\mu_0,\phi_0) = \xi(0,+\mu,\phi) + \frac{1}{4\pi}\int_0^1 \int_0^{2\pi} S(\mu,\phi;\mu',\phi')\xi(0,-\mu',\phi')\frac{d\mu'}{\mu'}d\phi'
\end{equation}

The detailed derivation is given in page.(93) of \cite{Chandrasekhar}. Now the principle of invariance in semi-infinite atmosphere is applicable only if the atmosphere shows the translational invariance. For such case the atmospheric emission $\boldsymbol{\beta}$ will be independent of optical depth $\tau$. For thermal emission this case can be realized by considering an isothermal atmosphere. Then the planck emission can be written as B(T), where $T_{\tau_0}=T_{\tau_1}=.....=T$. Thus, for an isothermal atmosphere, at $\tau=0$, the source function in eqn.\eqref{source function for thermal emission+scattering1} can be written using the boundary condition \eqref{boundary condition} as,
\begin{equation}\label{source function for emission+scattering2}
\xi(0,\mu,\phi) = B(T)+\frac{F}{16\pi}\int_{0}^1\int_{0}^{2\pi}p(\mu,\phi;\mu'',\phi'')S(\mu'',\phi'';\mu_0,\phi_0)\frac{d\mu''}{\mu''}d\phi''+\frac{1}{4}Fp(\mu,\phi;-\mu_0,\phi_0)
\end{equation}
Now putting eqn.\eqref{source function for emission+scattering2} in eqn.\eqref{Scattering function integral1} we will get the integral form of scattering function,
\begin{equation}\label{scattering function1}
\begin{split}
&(\frac{1}{\mu_0} +\frac{1}{\mu})S(\mu,\phi,\mu_0,\phi_0)\\ 
&=
4U(T)[1+\frac{1}{4\pi}\int_0^1 \int_0^{2\pi} S(\mu,\phi;\mu',\phi')\frac{d\mu'}{\mu'}d\phi']\\
&+ 
[p(\mu,\phi;-\mu_0,\phi_0) +\frac{1}{4\pi}\int_0^1 \int_0^{2\pi} S(\mu,\phi;\mu',\phi')p(-\mu',\phi';-\mu_0,\phi_0)\frac{d\mu'}{\mu'}d\phi']\\
&+\frac{1}{4\pi} [\int_0^1\int_{0}^{2\pi}p(\mu,\phi;\mu'',\phi'') \boldsymbol{S(\mu'',\phi'';\mu_0,\phi_0)} d\phi''\frac{d\mu''}{\mu''}\\
&+\frac{1}{4\pi}\int_0^1 \int_0^{2\pi} \int_0^1 \int_0^{2\pi} S(\mu,\phi;\mu',\phi')p(-\mu',\phi';\mu'',\phi'') \boldsymbol{S(\mu'',\phi'';\mu_0,\phi_0)} d\phi''\frac{d\mu''}{\mu''}\frac{d\mu'}{\mu'}d\phi']
\end{split}
\end{equation}
 where we define $U(T)=\frac{B(T)}{F}$.
This is the general form of scattering function, which is somewhat different from the equation derived in \cite{chandrasekhar1947radiative}.

\section{Explicit form of the scattering integral equation with different phase functions}\label{S with different p}
Atmospheric scattering can be characterized by the scattering phase function $p(\cos\Theta)$, where $\Theta$ is the angle between $(\mu,\phi)$ and $(\mu',\phi')$. For true absorption and scattering, phase function holds the following identity \citet{Chandrasekhar},
\begin{equation}
\frac{1}{4\pi}\int p(\cos(\Theta))d\Omega = \tilde{\omega_0}\leqslant 1
\end{equation}
 
The scattering function S is explicitly dependent on scattering phase function p as shown in section~\ref{integral form of scattering function}. Now, we will derive the reduced scattering integral equation for different phase functions.
\subsection{Isotropic phase function:}\label{scattering for isotropic}
The isotropic scattering phase function for \textit{semi-infinite homogeneous atmosphere} do not have any directional dependency and can be expressed as, $p(\cos\Theta) =p(\mu,\phi;\mu',\phi')= \tilde{\omega_0}$(Constant)$\leqslant1$, where $\tilde{\omega_0}$ is \textit{the single scattering albedo} \citet{Chandrasekhar}. For an atmosphere with both scattering as well as absorption \citet{sengupta2020optical}, $\tilde{\omega_0}$ can be defined as, 
\begin{equation}\label{single scattering albedo}
\tilde{\omega_0}=\frac{\sigma}{\chi}
\end{equation}
followed from \citet{domanus1974fundamental}.
 Note that, $\tilde{\omega_0}=1$ represents the \textit{ conservative case of pure scattering} \citet{Chandrasekhar}.

In view of axial symmetry, the source function S for isotropic phase function can be written using  eqn.\eqref{scattering function1} as,
\begin{equation}\label{scattering function isotropic}
\begin{split}
&(\frac{1}{\mu_0} +\frac{1}{\mu})S(\mu;\mu_0)\\ 
&=
4U(T)[1+\frac{1}{2}\int_0^1 S(\mu;\mu')\frac{d\mu'}{\mu'}]+ 
\tilde{\omega_0}[1 +\frac{1}{2}\int_0^1 S(\mu;\mu')\frac{d\mu'}{\mu'}][1+\frac{1}{2}\int_0^1   S(\mu_0;\mu'') \frac{d\mu''}{\mu''}]
\end{split}
\end{equation}

Those bracketed terms in right hand side must be the values of either $\mu$ or $\mu_0$ of the \textit{same function}. Let define the function as,
\begin{equation}\label{M-function1}
M(\mu)= 1+\frac{1}{2}\int_0^1 S(\mu;\mu')\frac{d\mu'}{\mu'}
\end{equation}
Thus, eqn.\eqref{scattering function isotropic} will reduce in terms of \textit{M-function},
\begin{equation}\label{scattering function isotropic1}
(\frac{1}{\mu_0} +\frac{1}{\mu})S(\mu;\mu_0)=4U(T)M(\mu) +\tilde{\omega_0}M(\mu)M(\mu_0)
\end{equation}
This M-function was derived by putting the expression of $S(\mu,\mu')$ (eqn.\eqref{scattering function isotropic1}) in eqn.\eqref{M-function1} as,
\begin{equation}\label{M-function2}
\therefore M(\mu)= 1+2U(T)M(\mu)\mu \log(1+\frac{1}{\mu})+ \frac{\tilde{\omega_0}}{2}\mu M(\mu)\int_0^1 \frac{M(\mu')}{\mu+\mu'}d\mu'
\end{equation}

Here we mention that the conservative case ($\tilde{\omega_0}=1$) will not need any special treatment. Just by replacing $\tilde{\omega_0}=1$ in eqns.\eqref{scattering function isotropic1} and \eqref{M-function2} we will get the exact forms for conservative case.

Finally, we determine diffusely reflected specific intensity from a thermally emitting atmosphere  $I(0,\mu.\mu_0)$ using eqn.\eqref{I0_1} for \textit{isotropic scattering} in terms of M-function as follows,
\begin{equation}\label{I(0)_isotropic}
\begin{split}
I(0,\mu,\mu_0)
&=\frac{F}{4\mu}\frac{\mu\mu_0}{\mu+\mu_0}[4U(T)+\tilde{\omega_0}M(\mu_0)]M(\mu)\\
 &= \frac{\mu_0}{\mu+\mu_0}B(T)M(\mu)+\frac{F}{4}\frac{\mu_0}{\mu+\mu_0}\tilde{\omega_0}M(\mu_0)M(\mu) 
\end{split}
\end{equation}
\subsection{Asymmetric scattering:}\label{scattering for linear}

The \textit{asymetric scattering phase function} can be written as, $p(\cos\Theta)=\tilde{\omega_0}(1+x\cos\Theta)$, where $\tilde{\omega_0}$ is single scattering albedo given in eqn.\eqref{single scattering albedo} and $x \in [0,1]$ is the asymmetric factor.
This type of phase function generally occurs for planetary illumination case introducing the assymetry \citet{Chandrasekhar}.

As,
\begin{equation}\label{cos theta}
\cos\Theta=\mu\mu'+\sqrt{(1-\mu^2)(1-\mu'^2)}\cos(\phi'-\phi)
\end{equation} 

then the explicit form of the phase function can be written as,
\begin{equation}\label{px1}
p(\mu,\phi;\mu',\phi')=\tilde{\omega_0}[1+x\mu\mu'+x\sqrt{(1-\mu^2)(1-\mu'^2)}\cos(\phi'-\phi)]
\end{equation}

 The scattering function here, will have the particular form \citet{Chandrasekhar},
\begin{equation}\label{sx1}
S(\mu,\phi;\mu',\phi')= \tilde{\omega_0}[S^{(0)}(\mu;\mu')+S^{(1)}(\mu;\mu')x\sqrt{(1-\mu^2)(1-\mu'^2)}\cos(\phi'-\phi)]
\end{equation}
To solve for the scattering function we need to use the following identity,
\begin{equation}\label{identity}
\begin{split}
\frac{1}{2\pi}\int_0^{2\pi}\cos m(\phi'-\phi_0)\cos n(\phi-\phi')d\phi'&=0;m\neq n\\
&=\frac{1}{2}\cos m(\phi-\phi_0);m=n\neq 0\\
&=1;m=n= 0
\end{split}
\end{equation}

Using eqns.\eqref{px1},\eqref{scattering function1} and \eqref{identity} we will deduce the form of $S(\mu,\phi;\mu_0,\phi_0)$ and comparing with eqn.\eqref{sx1} with $(\mu',\phi')$ replaced by $(\mu_0,\phi_0)$  we will get the form of $S^{(0)}$ as,
\begin{equation}\label{S^0 px case}
\begin{split}
(\frac{1}{\mu_0} +\frac{1}{\mu})S^{(0)}(\mu;\mu_0)
&=
\frac{4U(T)}{\tilde{\omega_0}}[1+\frac{\tilde{\omega_0}}{2}\int_0^1 S^{(0)}(\mu;\mu')\frac{d\mu'}{\mu'}]\\
&+[1+\frac{\tilde{\omega_0}}{2}\int_0^1S^{(0)}(\mu'',\mu_0)\frac{d\mu''}{\mu''}][1+\frac{\tilde{\omega_0}}{2}\int_0^1S^{(0)}(\mu',\mu)\frac{d\mu'}{\mu'}]\\
-&
x[\mu_0-\frac{\tilde{\omega_0}}{2}\int_0^1 S^{(0)}(\mu'',\mu_0)d\mu''][\mu-\frac{\tilde{\omega_0}}{2}\int_0^1 S^{(0)}(\mu',\mu)d\mu']
\end{split}
\end{equation}

We can write eqn.\eqref{S^0 px case} in closed form as follows,
\begin{equation}\label{S^0 px case1}
(\frac{1}{\mu_0} +\frac{1}{\mu})S^{(0)}_a(\mu;\mu_0)=\frac{4U(T)}{\tilde{\omega_0}}\psi_a(\mu) + \psi_a(\mu_0)\psi_a(\mu)-x\phi_a(\mu_0)\phi_a(\mu)
\end{equation}
where we define\footnote{ Hereinafter the subscript "a" denotes the \textit{asymmetric scattering case}.},
\begin{equation}\label{psi phi for px}
\begin{split}
\psi_a(\mu) = 1 + \frac{1}{2}\tilde{\omega_0}\int_0^1 S^{(0)}_a(\mu;\mu')\frac{d\mu'}{\mu'}\\
\phi_a(\mu) = \mu - \frac{1}{2}\tilde{\omega_0}\int_0^1 S^{(0)}_a(\mu;\mu') d\mu'
\end{split}
\end{equation}

Putting the expression of $S^{(0)}$ back into equation \eqref{psi phi for px} we will get the expressions of $\phi$ and $\psi$ as follows,
\begin{equation}\label{psi for xcos}
\begin{split}
\psi_a(\mu)
&= 1+2U(T)\psi_a(\mu)\mu\log(1+\frac{1}{\mu}) + \frac{\tilde{\omega_0}}{2}\mu\psi_a(\mu)\int_0^1\psi_a(\mu')\frac{d\mu'}{\mu+\mu'} - \frac{\tilde{\omega_0}}{2}x\mu\phi_a(\mu) \int_0^1\phi_a(\mu')\frac{d\mu'}{\mu+\mu'}
\end{split}
\end{equation}
and
\begin{equation}\label{phi for xcos}
\begin{split}
\phi_a(\mu) 
&= \mu - 2U(T)\psi_a(\mu)\mu\log(1+\frac{1}{\mu})-\frac{\tilde{\omega_0}}{2}\mu\psi_a(\mu)\int_0^1\psi_a(\mu')\frac{\mu'}{\mu+\mu'}d\mu' +\frac{\tilde{\omega_0}}{2}x\mu\phi_a(\mu)\int_0^1\phi_a(\mu')\frac{\mu'}{\mu+\mu'}d\mu'
\end{split}
\end{equation}

In the same way we can find the expression of $S^{(1)}(\mu,\mu_0)$ by comparing the $cos(\phi-\phi_0)$ term as follows \citet{Chandrasekhar},
\begin{equation}\label{S^1 px case}
\begin{split}
(\frac{1}{\mu_0} +\frac{1}{\mu})S^{(1)}(\mu;\mu_0)=&[1+\frac{\tilde{\omega_0}}{4}x\int_0^1S^{(1)}(\mu',\mu)(1-\mu'^2)\frac{d\mu'}{\mu'}]*[1+\frac{\tilde{\omega_0}}{4}x\int_0^1S^{(1)}(\mu'',\mu_0)(1-\mu''^2)\frac{d\mu''}{\mu''}]\\
=&
H^{(1)}(\mu)H^{(1)}(\mu_0)
\end{split}
\end{equation}
This $H^{(1)}(\mu)$ is defined as \citet{Chandrasekhar},
\begin{equation}\label{H^1 for px}
H^{(1)}(\mu)=1+\frac{\tilde{\omega_0}}{4}x \mu H^{(1)}(\mu) \int_0^1 \frac{H^{(1)}(\mu')}{\mu+\mu'} (1-\mu'^2) d\mu'
\end{equation}
Finally, we can put the values of $S^{(0)}$ and $S^{(1)}$ in eqn.\eqref{sx1} we will get,
\begin{equation}\label{sx2}
\begin{split}
&S_a(\mu,\phi;\mu_0,\phi_0)\\
&=\frac{\mu\mu_0}{\mu+\mu_0}4U(T)\psi_a(\mu)+\frac{\mu\mu_0}{\mu+\mu_0}\tilde{\omega_0}[(\psi_a(\mu)\psi_a(\mu_0)-x\phi_a(\mu)\phi_a(\mu_0))\\
&+H^{(1)}(\mu)H^{(1)}(\mu_0)x\sqrt{(1-\mu^2)(1-\mu_0^2)}\cos(\phi_0-\phi)]
\end{split}
\end{equation}
Thus, the diffusely reflected intensity from thermally emitting atmosphere in assymetric scattering can be determined using eqns. \eqref{sx2} and \eqref{I0_1} as,
\begin{equation}\label{I0 for px}
\begin{split}
I(0,\mu;\mu_0)& = \frac{F}{4\mu}S_a(\mu,\phi;\mu_0,\phi_0)\\
&=B(T)\frac{\mu_0}{\mu+\mu_0}\psi_a(\mu)+\frac{F}{4}\frac{\mu_0}{\mu+\mu_0}\tilde{\omega_0}[(\psi_a(\mu)\psi_a(\mu_0)-x\phi_a(\mu)\phi_a(\mu_0))\\
&+H^{(1)}(\mu)H^{(1)}(\mu_0)x\sqrt{(1-\mu^2)(1-\mu_0^2)}\cos(\phi_0-\phi)]
\end{split}
\end{equation}
\subsection{Rayleigh scattering  phase function:}\label{scattering for Rayleigh}

Rayleigh scattering is very important in astrophysical context. For planetary atmospheres it is the most dominant scattering process in the absence of Mie scattering \citet{madhusudhan2012analytic},\citet{kattawar1971flux}. For substellar-mass objects (e.g. T-dwarfs), multiple Rayleigh scattering shows polarization effect \citet{sengupta2009multiple}. In case of hot stars the neutral hydrogen and singly ionized helium shows Rayleigh scattering from the atmosphere \citet{fivsak2016rayleigh}.  Thus it is important to evaluate the scattering function for Rayleigh scattering case.

The scattering phase function can be written as  \cite{Chandrasekhar} ,
\begin{equation}
\begin{split}
p(\cos\Theta) &= \frac{3}{4}(1+\cos^2\Theta)\\
p(\mu,\phi;\mu',\phi') &= \frac{3}{8}[p^{(0)}(\mu,\mu') \\
&+ 4\mu\mu'\sqrt{(1-\mu^2)(1-\mu'^2)}\cos(\phi-\phi')\\
&+ (1-\mu^2)(1-\mu'^2)\cos2(\phi-\phi')]
\end{split}
\end{equation}
where, $p^{(0)}(\mu,\mu')$ satisfies the following relation,
\begin{equation}\label{p^0}
\begin{split}
p^{(0)}(\mu,\mu') =& \frac{1}{2\pi}\int_0^{2\pi}p(\mu,\phi;\mu',\phi') d\phi'\\
=& \frac{1}{3}(3-\mu^2)(3-\mu'^2) + \frac{8}{3}\mu^2\mu'^2
\end{split}
\end{equation}
We can express the scattering function by comparing with the phase function as,
\begin{equation}\label{s rayleigh}
\begin{split}
S(\mu,\phi;\mu',\phi') &= \frac{3}{8}[S^{(0)}(\mu,\mu')\\
&+ S^{(1)}(\mu,\mu')4\mu\mu'\sqrt{(1-\mu^2)(1-\mu'^2)}\cos(\phi-\phi')\\
&+ S^{(2)}(\mu,\mu')(1-\mu^2)(1-\mu'^2)\cos2(\phi-\phi')]
\end{split}
\end{equation}
Following the same procedure as we have done before we can derive the expressions of $S^{(0)}(\mu,\mu_0)$,$S^{(1)}(\mu,\mu_0)$ and $S^{(2)}(\mu,\mu_0)$.

\begin{equation}\label{s0 rayleigh1}
\begin{split}
&(\frac{1}{\mu_0} +\frac{1}{\mu})S^{(0)}(\mu,\mu_0)\\
&=\frac{8}{3}4U(T)[1+\frac{1}{2}\int_0^1  S^{(0)}(\mu;\mu')\frac{d\mu'}{\mu'}]  \\ 
&+\frac{1}{3}[3-\mu^2 + \frac{3}{16}\int_0^1(3-\mu'^2)S^{(0)}(\mu,\mu')\frac{d\mu'}{\mu'}]*[3-\mu_0^2 + \frac{3}{16}\int_0^1(3-\mu''^2)S^{(0)}(\mu_0,\mu'')\frac{d\mu''}{\mu''}]\\
&+ \frac{8}{3}[\mu^2+\frac{3}{16}\int_0^1 \mu'^2S^{(0)}(\mu,\mu')\frac{d\mu'}{\mu'}]*[\mu_0^2+\frac{3}{16}\int_0^1 \mu''^2S^{(0)}(\mu_0,\mu'')\frac{d\mu''}{\mu''}]
\end{split}
\end{equation}

We define the following terms\footnote{Hereinafter the sbscript "R" stands for Rayleigh scattering},
\begin{equation}
\begin{split}\label{phi psi gamma}
\psi_R(\mu)&= 3-\mu^2 + \frac{3}{16}\int_0^1(3-\mu'^2)S^{(0)}_R(\mu,\mu')\frac{d\mu'}{\mu'}\\
\phi_R(\mu) &= \mu^2 + \frac{3}{16}\int_0^1 \mu'^2 S^{(0)}_R(\mu,\mu')\frac{d\mu'}{\mu'}\\
\gamma_R(\mu) &= 1 + \frac{3}{16}\int_0^1S^{(0)}_R(\mu,\mu')\frac{d\mu'}{\mu'}
\end{split}
\end{equation}
Now eqn.\eqref{s0 rayleigh1} can be expressed as,
\begin{equation}\label{s0 rayleigh2}
\therefore (\frac{1}{\mu_0} +\frac{1}{\mu})S^{(0)}_R(\mu,\mu_0)=\frac{32}{3}U(T)\gamma_R(\mu)+ \frac{1}{3}\psi_R(\mu)\psi_R(\mu_0)+\frac{8}{3}\phi_R(\mu)\phi_R(\mu_0)
\end{equation}
Now putting eqn.\eqref{s0 rayleigh2} in eqn.\eqref{phi psi gamma} we can get the explicit forms for $\phi_R$,$\psi_R$ and $\gamma_R$ as follows,
\begin{equation}\label{gamma for rayleigh}
\gamma_R(\mu)=
1+2U(T)\gamma_R(\mu)\mu\log(1+\frac{1}{\mu})+ \frac{1}{16}\psi_R(\mu)\mu\int_0^1\frac{d\mu'}{\mu+\mu'}\psi_R(\mu') + \frac{1}{2}\phi_R(\mu)\mu\int_0^1\frac{d\mu'}{\mu+\mu'}\phi_R(\mu')
\end{equation}
and
\begin{equation}\label{psi for rayleigh}
\begin{split}
\psi_R(\mu)&= (3-\mu^2)[1 + 2U(T)\gamma_R(\mu)\mu\log(1+\frac{1}{\mu})]+\frac{3}{4}U(T)\gamma_R(\mu)\mu[\mu -\frac{1}{2}] +
\frac{1}{16}\mu\psi_R(\mu)\int_0^1\frac{3-\mu'^2}{\mu+\mu'}\psi_R(\mu')d\mu'\\
&+ \frac{1}{2}\mu\phi_R(\mu)\int_0^1\frac{3-\mu'^2}{\mu+\mu'}\phi_R(\mu')d\mu'
\end{split}
\end{equation}
finally
\begin{equation}\label{phi for rayleigh}
\begin{split}
\phi_R(\mu)&=
\mu^2[1+2U(T)\gamma_R(\mu)\mu\log(1+\frac{1}{\mu})] + \frac{3}{4}U(T)\mu\gamma_R(\mu)[\frac{1}{2}-\mu ] + \frac{1}{16}\mu\psi_R(\mu)\int_0^1\frac{\mu'^2}{\mu+\mu'}\psi_R(\mu')d\mu'\\
&+ \frac{1}{2}\mu\phi_R(\mu)\int_0^1\frac{\mu'^2}{\mu+\mu'}\phi_R(\mu')d\mu'
\end{split}
\end{equation}

The remaining expressions for $S^{(1)}(\mu,\mu')$ and $S^{(2)}(\mu,\mu')$ can be found by comparing the co-efficients of $\cos(\phi-\phi_0)$ and $\cos2(\phi-\phi_0)$ respectively (detailed derivation given in \citet{Chandrasekhar}). We write them down here,
\begin{equation}\label{s1 for rayleigh1}
\begin{split}
(\frac{1}{\mu} + \frac{1}{\mu_0})S^{(1)}(\mu,\mu_0) &= [1+\frac{3}{8}\int_0^1\mu''^2(1-\mu''^2)S^{(1)}(\mu'',\mu_0)\frac{d\mu''}{\mu''}]\\
& *[1+\frac{3}{8}\int_0^1\mu'^2(1-\mu'^2)S^{(1)}(\mu,\mu')\frac{d\mu'}{\mu'}]
\end{split}
\end{equation}
and,
\begin{equation}\label{s2 for rayleigh1}
\begin{split}
(\frac{1}{\mu} + \frac{1}{\mu_0})S^{(2)}(\mu,\mu_0) &= [1+\frac{3}{32}\int_0^1(1-\mu''^2)^2S^{(2)}(\mu'',\mu_0)\frac{d\mu''}{\mu''}]\\
&X[1+\frac{3}{32}\int_0^1(1-\mu'^2)^2S^{(2)}(\mu,\mu')\frac{d\mu'}{\mu'}]
\end{split}
\end{equation}
Now, eqns. \eqref{s1 for rayleigh1} and \eqref{s2 for rayleigh1} can be expressed in terms of $H^{(1)}(\mu)$ and $H^{(2)}(\mu)$ as follows \citet{Chandrasekhar},
\begin{equation}
\begin{split}\label{s1 and s2 for rayleigh}
(\frac{1}{\mu} + \frac{1}{\mu_0})S^{(1)}(\mu,\mu_0)=
H^{(1)}(\mu_0)H^{(1)}(\mu)\\
(\frac{1}{\mu} + \frac{1}{\mu_0})S^{(2)}(\mu,\mu_0)=
H^{(2)}(\mu_0)H^{(2)}(\mu)
\end{split}
\end{equation}
Where $H^{(1)}(\mu)$ and $H^{(2)}(\mu)$ can be expressed as,
\begin{equation}\label{H^1 rayleigh}
H^{(1)}(\mu) = 1+\frac{3}{8}\mu H^{(1)}(\mu)\int_0^1 \frac{\mu'^2(1-\mu'^2)}{\mu+\mu'}H^{(1)}(\mu')d\mu'
\end{equation}
\begin{equation}\label{H^2 rayleigh}
H^{(2)}(\mu) =1+\frac{3}{32}\mu H^{(2)}(\mu)\int_0^1\frac{(1-\mu'^2)^2}{\mu+\mu'}H^{(2)}(\mu')d\mu'
\end{equation}

Now the full equation of scattering function and intensity at the layer of $\tau=0$ for Rayleigh scattering can be expressed as follows,
\begin{equation}\label{s rayleigh1}
\begin{split}
(\frac{1}{\mu} + \frac{1}{\mu_0})S_R(\mu,\phi;\mu_0,\phi_0) &= \frac{3}{8}[\frac{32}{3}U(T)\gamma_R(\mu)+ \frac{1}{3}\psi_R(\mu)\psi_R(\mu_0)+\frac{8}{3}\phi_R(\mu)\phi_R(\mu_0)\\
&- H^{(1)}(\mu)H^{(1)}(\mu_0)4\mu\mu_0\sqrt{(1-\mu^2)(1-\mu_0^2)}\cos(\phi-\phi_0)\\
&+ H^{(2)}(\mu)H^{(2)}(\mu_0)(1-\mu^2)(1-\mu_0^2)\cos2(\phi-\phi_0)]
\end{split}
\end{equation}
and
\begin{equation}\label{I_(0) Rayleigh}
\begin{split}
I(0,\mu;\mu_0) &= \frac{F}{4\mu}S_R(\mu,\phi;\mu_0,\phi_0)\\
&= \frac{\mu_0}{\mu+\mu_0}B(T)\gamma_R(\mu)\\
&+ \frac{3F}{32}\frac{\mu_0}{\mu+\mu_0}[\frac{1}{3}\psi_R(\mu)\psi_R(\mu_0)+\frac{8}{3}\phi_R(\mu)\phi_R(\mu_0)\\
&- H^{(1)}(\mu)H^{(1)}(\mu_0)4\mu\mu_0\sqrt{(1-\mu^2)(1-\mu_0^2)}\cos(\phi-\phi_0)\\
&+ H^{(2)}(\mu)H^{(2)}(\mu_0)(1-\mu^2)(1-\mu_0^2)\cos2(\phi-\phi_0)]
\end{split}
\end{equation}

\subsection{Scattering function for the phase function $p(\cos\Theta)=\tilde{\omega_0}+\tilde{\omega_1}P_1(\cos\Theta)+\tilde{\omega_2}P_2(\cos\Theta)$:}\label{scattering for legendre}
In this section we use the phase function expanded in terms of Legendre polynomials upto order 2.
\begin{equation}\label{phase function legendre}
\begin{split}
p(\cos\Theta)=&\sum_{m=0}^{2}\tilde{\omega_m}P_m(\cos \Theta)\\
=&
\tilde{\omega_0}+\tilde{\omega_1}P_1(\cos\Theta)+\tilde{\omega_2}P_2(\cos\Theta)
\end{split}
\end{equation}

where $\tilde{\omega_0},\tilde{\omega_1},\tilde{\omega_2}$ are constants and $P_1,P_2$ are Legendre polynomials of $\cos\Theta$ given in eqn.\eqref{cos theta}. The analytic solutions for semi-infinite atmosphere using this phase function in case of diffuse reflection only have been obtained in \citet{horak1961diffuse}. In this paper we introduce the thermal emission effect with it. Note that at different values of $\tilde{\omega_0}$(\textit{single scattering albedo}), $\tilde{\omega_1}$ and $\tilde{\omega_2}$ this phase function will reduce into the previous phase functions. If $\tilde{\omega_1}=\tilde{\omega_2}=0$ then it is isotropic scattering, if $\tilde{\omega_2}=0,\tilde{\omega}_1=x\tilde{\omega}_0$, then it is asymmetric scattering and for rayleigh scattering $\tilde{\omega_0}=1,\tilde{\omega_1}=0,\tilde{\omega_2}=\frac{1}{2}$. For an extensive discussion of different phase functions produced due to the choices of different set of $\tilde{\omega_0},\tilde{\omega_1},\tilde{\omega_2}$ values we refer \citet{bhatia1983multiple}. Thus, we can say this is the most general phase function till now discussed.

Considering \textit{single scattering albedo} $\tilde{\omega_0}\leqslant1$ i.e. non-conservative case the phase function can be written explicitly as \citet{horak1961diffuse},
\begin{equation}\label{phase function legendre1}
\begin{split}
p(\mu,\phi;\mu',\phi') &= p^{(0)}(\mu,\mu') \\
&+ (\tilde{\omega_1}+3\tilde{\omega_2}\mu\mu')\sqrt{(1-\mu^2)(1-\mu'^2)}\cos(\phi-\phi')\\
&+ \frac{3}{4}\tilde{\omega_2}(1-\mu^2)(1-\mu'^2)\cos2(\phi-\phi')
\end{split}
\end{equation}
Following, \eqref{p^0}, $p^{(0)}(\mu,\mu')$ can be written as,
\begin{equation}\label{p^0 for legendre}
\begin{split}
p^{(0)}(\mu,\mu') =&\frac{3\tilde{\omega_2}}{4}\frac{1}{\zeta}[\zeta-\mu^2)(\zeta-\mu'^2)+\frac{4\tilde{\omega_0}}{\tilde{\omega_2}}(\mu\mu')^2] + \tilde{\omega_1}\mu\mu'
\end{split}
\end{equation}
where $\zeta=\frac{4\tilde{\omega_0}+\tilde{\omega_2}}{3\tilde{\omega_2}}$.
The scattering function can be written as before,
\begin{equation}\label{scattering function legendre}
\begin{split}
S(\mu,\phi;\mu_0,\phi_0) &= S^{(0)}(\mu,\mu_0) \\
&+ S^{(1)}(\mu,\mu_0)\sqrt{(1-\mu^2)(1-\mu_0^2)}\cos(\phi-\phi_0)\\
&+ S^{(2)}(\mu,\mu_0)(1-\mu^2)(1-\mu_0^2)\cos2(\phi-\phi_0)
\end{split}
\end{equation}

It has been shown in section~\ref{scattering for linear} and \ref{scattering for Rayleigh} that $S^{(1)}(\mu,\mu_0)$ or $S^{(2)}(\mu,\mu_0)$ are not effected by thermal emission $B(T_\tau)$ and only $S^{(0)}(\mu,\mu')$ is effected. Thus, here we show the calculations $S^{(0)}(\mu,\mu')$ only and write the expressions of $S^{(1)}(\mu,\mu'),S^{(2)}(\mu,\mu')$ as given in \citet{horak1961diffuse}.

Now putting phase function eqn.\eqref{phase function legendre} in scattering function expression \eqref{scattering function1} and comparing with eqn.\eqref{scattering function legendre} we will get\footnote{Hereinafter the subscript "l" stands for the scattering due to phase function $p(\cos\Theta)=\tilde{\omega_0}+\tilde{\omega_1}P_1(\cos\Theta)+\tilde{\omega_2}P_2(\cos\Theta)$},

\begin{equation}\label{S^0 for legendre}
(\frac{1}{\mu}+\frac{1}{\mu_0})S^{(0)}_l(\mu,\mu_0) = 4U(T)\gamma_l(\mu) -\tilde{\omega_1}\eta_l(\mu)\eta_l(\mu_0) + \frac{3\tilde{\omega_0}}{\zeta}\phi_l(\mu)\phi_l(\mu_0) + \frac{3\tilde{\omega_2}}{4\zeta}\psi_l(\mu)\psi_l(\mu_0)
\end{equation}
where, 
\begin{equation}\label{gamma for legendre}
\gamma_l(\mu) = 1 + \frac{1}{2}\int_0^1S^{(0)}_l(\mu,\mu')\frac{d\mu'}{\mu'}
\end{equation}
\begin{equation}\label{eta for legendre}
\eta_l(\mu) = \mu - \frac{1}{2}\int_0^1S^{(0)}_l(\mu,\mu')d\mu'
\end{equation}
\begin{equation}\label{phi for legendre}
\phi_l(\mu) = \mu^2 + \frac{1}{2}\int_0^1S^{(0)}_l(\mu,\mu') \mu'd\mu'
\end{equation}
\begin{equation}\label{psi for legendre}
\psi_l(\mu) = (\zeta-\mu^2) + \frac{1}{2}\int_0^1S^{(0)}_l(\mu,\mu')(\zeta-\mu'^2)\frac{d\mu'}{\mu'}
\end{equation}
To write the explicit expressions of $\gamma_l,\eta_l,\phi_l$ and $\psi_l$ we put the expression \eqref{S^0 for legendre} in equations \eqref{gamma for legendre}-\eqref{psi for legendre} and get the following expressions,
\begin{equation}\label{gamma for legendre1}
\begin{split}
\gamma_l(\mu)
=& 1+2U(T)\mu\gamma_l(\mu)\log(1+\frac{1}{\mu}) -\frac{\mu}{2}\tilde{\omega_1}\eta_l(\mu)\int_0^1\frac{\eta_l(\mu')}{\mu+\mu'}d\mu'
+\frac{\mu}{2}\frac{3\tilde{\omega_0}}{\zeta}\phi_l(\mu)\int_0^1\frac{\phi_l(\mu')}{\mu+\mu'}d\mu'\\
&+\frac{\mu}{2}\frac{3\tilde{\omega_2}}{4\zeta}\psi_l(\mu)\int_0^1\frac{\psi_l(\mu')}{\mu+\mu'}d\mu'
\end{split}
\end{equation}

\begin{equation}\label{eta for legendre1}
\begin{split}
\eta_l(\mu)
=&\mu - 2U(T)\gamma_l(\mu)\mu[1-\mu\log(1+\frac{1}{\mu})]+\frac{\mu}{2}\tilde{\omega_1}\eta_l(\mu)\int_0^1\frac{\eta_l(\mu')}{\mu+\mu'}\mu'd\mu'\\
&-\frac{\mu}{2}\frac{3\tilde{\omega_0}}{\zeta}\phi_l(\mu)\int_0^1\frac{\phi_l(\mu')}{\mu+\mu'}\mu'd\mu'-\frac{\mu}{2}\frac{3\tilde{\omega_2}}{4\zeta}\psi_l(\mu)\int_0^1\frac{\psi_l(\mu')}{\mu+\mu'}\mu'd\mu'
\end{split}
\end{equation}

\begin{equation}\label{phi for legendre1}
\begin{split}
\phi_l(\mu)
=& \mu^2+2U(T)\gamma_l(\mu)\mu[\frac{1}{2}-\mu+\mu^2\log(1+\frac{1}{\mu})]-\frac{\mu}{2}\tilde{\omega_1}\eta_l(\mu)\int_0^1\frac{\eta_l(\mu')}{\mu+\mu'}\mu'^2d\mu'\\
&+\frac{\mu}{2}\frac{3\tilde{\omega_0}}{\zeta}\phi_l(\mu)\int_0^1\frac{\phi_l(\mu')}{\mu+\mu'}\mu'^2d\mu'
+\frac{\mu}{2}\frac{3\tilde{\omega_2}}{4\zeta}\psi_l(\mu)\int_0^1\frac{\psi_l(\mu')}{\mu+\mu'}\mu'^2d\mu'
\end{split}
\end{equation}

\begin{equation}\label{psi for legendre1}
\begin{split}
\psi_l(\mu)
=& (\zeta-\mu^2) + 2U(T)\mu\gamma_l(\mu)[(\zeta-\mu^2)\log(1+\frac{1}{\mu})+\mu-\frac{1}{2}] -\frac{\mu}{2}\tilde{\omega_1}\eta_l(\mu)\int_0^1\frac{\eta_l(\mu')}{\mu+\mu'}(\zeta-\mu'^2)d\mu'\\
&+\frac{\mu}{2}\frac{3\tilde{\omega_0}}{\zeta}\phi_l(\mu)\int_0^1\frac{\phi_l(\mu')}{\mu+\mu'}(\zeta-\mu'^2)d\mu'
+\frac{\mu}{2}\frac{3\tilde{\omega_2}}{4\zeta}\psi_l(\mu)\int_0^1\frac{\psi_l(\mu')}{\mu+\mu'}(\zeta-\mu'^2)d\mu'
\end{split}
\end{equation}

Finally the expressions of $S^{(1)}(\mu,\mu_0),S^{(2)}(\mu,\mu_0)$ given in \citet{horak1961diffuse} as,
\begin{equation}\label{S^1 for legendre}
\begin{split}
(\frac{1}{\mu}+\frac{1}{\mu_0})S^{(1)}(\mu,\mu_0) =& \omega_1\theta(\mu)\theta(\mu_0) - 3\omega_2\sigma(\mu)\sigma(\mu_0)\\
=&[\tilde{\omega_1}(1+l\mu)(1+l\mu_0)-3\tilde{\omega_2}m^2\mu\mu_0]H^{(1)}(\mu)H^{(1)}(\mu_0)
\end{split}
\end{equation}
where l and m are the same as given in \citet{horak1961diffuse}
and,
\begin{equation}\label{S^2 for legendre}
(\frac{1}{\mu}+\frac{1}{\mu_0})S^{(2)}(\mu,\mu_0) = \frac{3\tilde{\omega_2}}{4}H^{(2)}(\mu)H^{(2)}(\mu_0)
\end{equation}
The H-functions here defined as,
\begin{equation}\label{H-function legendre}
H^{(i)}(\mu) = 1+\mu H^{(i)}(\mu)\int_0^1\frac{\Psi^{(i)}(\mu')}{\mu+\mu'}H^{(i)}(\mu')d\mu'\hspace{1cm}i=1,2
\end{equation}
where,
\begin{eqnarray}\label{variables}
\Psi^{(i)}(\mu) = a^{(i)} + b^{(i)}\mu^2 + c^{(i)}\mu^4\hspace{3cm}\\
a^{(1)}=\frac{\tilde{\omega_1}}{4};\hspace{0.5cm} b^{(1)}=\frac{1}{4}[\tilde{\omega_2}(3-\tilde{\omega_1})-\tilde{\omega_1}];\hspace{0.5cm}c^{(1)} = \frac{\tilde{\omega_2}}{4}(\tilde{\omega_1}-3)\\
a^{(2)}=c^{(2)}=\frac{3\tilde{\omega_2}}{16};\hspace{2cm}b^{(2)} = -\frac{3\omega_2}{8}\hspace{2cm}
\end{eqnarray}

Then the total scattering function will be,
\begin{equation}\label{scattering function legendre1}
\begin{split}
&S_l(\mu,\phi;\mu_0,\phi_0)\\ &= \frac{\mu\mu_0}{\mu+\mu_0}\{4U(T)\gamma_l(\mu) -\tilde{\omega_1}\eta_l(\mu)\eta_l(\mu_0) + \frac{3\tilde{\omega_0}}{\zeta}\phi_l(\mu)\phi_l(\mu_0) + \frac{3\tilde{\omega_2}}{4\zeta}\psi_l(\mu)\psi_l(\mu_0)] \\
&+ [\tilde{\omega_1}(1+l\mu)(1+l\mu_0)-3\tilde{\omega_2}m^2\mu\mu_0]H^{(1)}(\mu)H^{(1)}(\mu_0)\sqrt{(1-\mu^2)(1-\mu_0^2)}\cos(\phi-\phi_0)\\
&+ \frac{3\tilde{\omega_2}}{4}H^{(2)}(\mu)H^{(2)}(\mu_0)(1-\mu^2)(1-\mu_0^2)\cos2(\phi-\phi_0)\}
\end{split}
\end{equation}
Now the intensity $I(0,\mu;\mu_0)$ here can be expressed as,
\begin{equation}\label{I^(0) legendre}
\begin{split}
I(0,\mu;\mu_0) 
=&\frac{\mu_0}{\mu+\mu_0}B(T)\gamma_l(\mu)\\
&+\frac{\mu_0}{\mu+\mu_0}\frac{F}{4}[-\tilde{\omega_1}\eta_l(\mu)\eta_l(\mu_0) + \frac{3\tilde{\omega_0}}{\zeta}\phi_l(\mu)\phi_l(\mu_0) + \frac{3\tilde{\omega_2}}{4\zeta}\psi_l(\mu)\psi_l(\mu_0)\\
&+\{\tilde{\omega_1}(1+l\mu)(1+l\mu_0)-3\tilde{\omega_2}m^2\mu\mu_0\}H^{(1)}(\mu)H^{(1)}(\mu_0)\sqrt{(1-\mu^2)(1-\mu_0^2)}\cos(\phi-\phi_0)\\
&+\frac{3\tilde{\omega_2}}{4}H^{(2)}(\mu)H^{(2)}(\mu_0)(1-\mu^2)(1-\mu_0^2)\cos2(\phi-\phi_0)]
\end{split}
\end{equation}


\section{Comparison of our general model with Chandrasekhar's diffusion scattering model}\label{comparison with chandrasekhar}

We introduced thermal emission effect in Chandrasekhar's semi-infinite diffused reflection problem. One can expect that all of our present results will reduce into those of Chandrasekhar's results for negligeble thermal emission. Here we consider the case $B(T)<<F$ which can also be considered as $U(T)\to 0$ to reduce our results in the limit of scattering only atmosphere and compare with previous results as given in \citet{Chandrasekhar},\citet{horak1950diffuse} and \citet{horak1961diffuse}
\begin{enumerate}
\item \textit{The Transfer equation \eqref{emission with diffusion approximation} will be:}
\begin{equation}\label{only diffusion approximation}
\begin{split}
\mu\frac{dI(\tau,\mu,\phi)}{d\tau} =& I(\tau,\mu,\phi)-\frac{1}{4\pi}\int_{-1}^1\int_{0}^{2\pi}p(\mu,\phi;\mu',\phi')I(\tau,\mu',\phi')d\mu'd\phi'\\
&-\frac{1}{4}Fe^{-\tau/\mu_0}p(\mu,\phi;-\mu_0,\phi_0)
\end{split}
\end{equation}
Same as, eqn. (126) \citet[pg. 22 ]{Chandrasekhar}
\item \textit{The integral equation \eqref{scattering function1} for Scattering function S will be:}
\begin{equation}\label{scattering function1 without emission}
\begin{split}
(\frac{1}{\mu_0} +\frac{1}{\mu})&S(\mu,\phi,\mu_0,\phi_0)=p(\mu,\phi;-\mu_0,\phi_0) \\ 
&+\frac{1}{4\pi}\int_0^1 \int_0^{2\pi} S(\mu,\phi;\mu',\phi')p(-\mu',\phi';-\mu_0,\phi_0)\frac{d\mu'}{\mu'}d\phi'\\
&+\frac{1}{4\pi} \int_0^1\int_{0}^{2\pi}p(\mu,\phi;\mu'',\phi'') \boldsymbol{S(\mu'',\phi'';\mu_0,\phi_0)} d\phi''\frac{d\mu''}{\mu''}\\
&+\frac{1}{16\pi^2}\int_0^1 \int_0^{2\pi} \int_0^1 \int_0^{2\pi} S(\mu,\phi;\mu',\phi')p(-\mu',\phi';\mu'',\phi'') \boldsymbol{S(\mu'',\phi'';\mu_0,\phi_0)} d\phi''\frac{d\mu''}{\mu''}\frac{d\mu'}{\mu'}d\phi'
\end{split}
\end{equation}
Same as, eqn. (28) \citet[pg. 94]{Chandrasekhar}
\item The equations for isotropic scattering, \begin{enumerate}
\item \textit{The equation \eqref{scattering function isotropic1} of scattering function for isotropic case will be:}
\begin{equation}\label{scattering function isotropic1 without emission}
(\frac{1}{\mu_0} +\frac{1}{\mu})S(\mu;\mu_0)=\tilde{\omega_0}M(\mu)M(\mu_0)
\end{equation}
\item\textit{M-function defined in eqn.\eqref{M-function2} will be:}
\begin{equation}\label{M-function2 without emission}
\therefore M(\mu)= 1+\frac{\tilde{\omega_0}}{2}\mu M(\mu)\int_0^1 \frac{M(\mu')}{\mu+\mu'}d\mu'
\end{equation}
\item \textit{The intensity equation \eqref{I(0)_isotropic} will be:}
\begin{equation}\label{I(0)_isotropic without emission}
I(0,\mu,\mu_0)=\frac{F}{4}\frac{\mu_0}{\mu+\mu_0}\tilde{\omega_0}M(\mu_0)M(\mu) 
\end{equation}
\end{enumerate}
This reduced form of M-function is equivalent to \textit{Chandrasekhar's H-function} given in \citet[pg. 90, eqn.(42)]{Chandrasekhar}. The scattering function and intensity expressions are also equivalent to that given in \citet{Chandrasekhar} and \citet{horak1950diffuse} only the M-functions replaced by H-function.

\item The equations for the asymmetric scattering phase function, $\tilde{\omega_0}(1+x\cos\theta)$:
\begin{enumerate}
\item \textit{The scattering function of zeroth order $S_a^{(0)}$ defined in equation \eqref{S^0 px case1} will be:}
\begin{equation}\label{S^0 px case1 without emission}
(\frac{1}{\mu_0} +\frac{1}{\mu})S_a^{(0)}(\mu;\mu_0)= \psi_a(\mu_0)\psi_a(\mu)-x\phi_a(\mu_0)\phi_a(\mu)
\end{equation}
\item \textit{The $\psi_a(\mu)$ function defined in equation \eqref{psi for xcos} will be:}
\begin{equation}\label{psi for xcos without emission}
\begin{split}
\psi_a(\mu)
&= 1+ \frac{\tilde{\omega_0}}{2}\mu\psi_a(\mu)\int_0^1\psi_a(\mu')\frac{d\mu'}{\mu+\mu'} - \frac{\tilde{\omega_0}}{2}x\mu\phi_a(\mu) \int_0^1\phi_a(\mu')\frac{d\mu'}{\mu+\mu'}
\end{split}
\end{equation}
\item \textit{The $\phi(\mu)$ function defined in equation \eqref{phi for xcos} will be:}
\begin{equation}\label{phi for xcos without emission}
\begin{split}
\phi_a(\mu) 
&= \mu -\frac{\tilde{\omega_0}}{2}\mu\psi_a(\mu)\int_0^1\psi_a(\mu')\frac{\mu'}{\mu+\mu'}d\mu' +\frac{\tilde{\omega_0}}{2}x\mu\phi_a(\mu)\int_0^1\phi_a(\mu')\frac{\mu'}{\mu+\mu'}d\mu'
\end{split}
\end{equation}
\item \textit{The intensity equation \eqref{I0 for px} will be:}
\begin{equation}\label{I0 for px without emission}
\begin{split}
I(0,\mu;\mu_0)
&=\frac{F}{4}\frac{\mu_0}{\mu+\mu_0}\tilde{\omega_0}[(\psi_a(\mu)\psi_a(\mu_0)-x\phi_a(\mu)\phi_a(\mu_0))\\
&+H^{(1)}(\mu)H^{(1)}(\mu_0)x\sqrt{(1-\mu^2)(1-\mu_0^2)}\cos(\phi_0-\phi)]
\end{split}
\end{equation}
\end{enumerate}
The expressions of $S^{(0)},\psi_a$ and $\phi_a$ are all same as given in \citet[pg. 100, eqns.(59),(61) and (62)]{Chandrasekhar}. The intensity is exactly same as given in \citet{horak1950diffuse}.

\item Reduced equations for the Rayleigh scattering phase function, $\frac{3}{4}(1+\cos^2\theta)$:
\begin{enumerate}
\item \textit{The scattering function of zeroth order $S_R^{(0)}$ defined in equation \eqref{s0 rayleigh2} will change as:}
\begin{equation}\label{s0 rayleigh2 without emission:}
\therefore (\frac{1}{\mu_0} +\frac{1}{\mu})S^{(0)}_R(\mu,\mu_0)= \frac{1}{3}\psi_R(\mu)\psi_R(\mu_0)+\frac{8}{3}\phi_R(\mu)\phi_R(\mu_0)
\end{equation}
\item \textit{The $\psi_R(\mu)$ defined in equation \eqref{psi for rayleigh} will change as:}
\begin{equation}\label{psi for rayleigh without emission}
\begin{split}
\psi_R(\mu)&= (3-\mu^2) +
\frac{1}{16}\mu\psi_R(\mu)\int_0^1\frac{3-\mu'^2}{\mu+\mu'}\psi_R(\mu')d\mu'+ \frac{1}{2}\mu\phi_R(\mu)\int_0^1\frac{3-\mu'^2}{\mu+\mu'}\phi_R(\mu')d\mu'
\end{split}
\end{equation}
\item \textit{The functional form of $\phi_R(\mu)$ defined in equation \eqref{phi for rayleigh} will be,}

\begin{equation}\label{phi for rayleigh without emission}
\phi_R(\mu)=
\mu^2 + \frac{1}{16}\mu\psi_R(\mu)\int_0^1\frac{\mu'^2}{\mu+\mu'}\psi_R(\mu')d\mu'+ \frac{1}{2}\mu\phi_R(\mu)\int_0^1\frac{\mu'^2}{\mu+\mu'}\phi_R(\mu')d\mu'
\end{equation}
\item \textit{The intensity equation \eqref{I_(0) Rayleigh} will be modified as:} 
\begin{equation}\label{I_(0) Rayleigh without emission}
\begin{split}
I(0,\mu;\mu_0) 
&=  \frac{3F}{32}\frac{\mu_0}{\mu+\mu_0}[\frac{1}{3}\psi_R(\mu)\psi_R(\mu_0)+\frac{8}{3}\phi_R(\mu)\phi_R(\mu_0)\\
&- H^{(1)}(\mu)H^{(1)}(\mu_0)4\mu\mu_0\sqrt{(1-\mu^2)(1-\mu_0^2)}\cos(\phi_0-\phi)\\
&+ H^{(2)}(\mu)H^{(2)}(\mu_0)(1-\mu^2)(1-\mu_0^2)\cos2(\phi_0-\phi)]
\end{split}
\end{equation}

\end{enumerate}
The reduced expressions of $S^{(0)}_R,\psi_R$ and $\phi_R$ for Rayleigh scattering are same with those given in page. 102, eqns. (77),(79) and (80) and $I(0,\mu;\mu_0)$ as given in page. (143) of \citet{Chandrasekhar}.

\item Reduced equations for the phase function, $\tilde{\omega_0}+\tilde{\omega_1}P_1(\cos\Theta)+\tilde{\omega_2}P_2(\cos\Theta)$
\begin{enumerate}
\item The functional form of $S_l^{(0)}(\mu,\mu_0)$ will be,
\begin{equation}\label{S^0 for legendre without emission}
(\frac{1}{\mu}+\frac{1}{\mu_0})S_l^{(0)}(\mu,\mu_0) = -\tilde{\omega_1}\eta_l(\mu)\eta_l(\mu_0) + \frac{3\tilde{\omega_0}}{\zeta}\phi_l(\mu)\phi_l(\mu_0) + \frac{3\tilde{\omega_2}}{4\zeta}\psi_l(\mu)\psi_l(\mu_0)
\end{equation}
\item The functional form of $\eta_l(\mu)$ will be,
\begin{equation}\label{eta for legendre1 without emission}
\begin{split}
\eta_l(\mu)
=&\mu +\frac{\mu}{2}\tilde{\omega_1}\eta_l(\mu)\int_0^1\frac{\eta_l(\mu')}{\mu+\mu'}\mu'd\mu'-\frac{\mu}{2}\frac{3\tilde{\omega_0}}{\zeta}\phi_l(\mu)\int_0^1\frac{\phi_l(\mu')}{\mu+\mu'}\mu'd\mu'\\
&-\frac{\mu}{2}\frac{3\tilde{\omega_2}}{4\zeta}\psi_l(\mu)\int_0^1\frac{\psi_l(\mu')}{\mu+\mu'}\mu'd\mu'
\end{split}
\end{equation}

\item The functional form of $\phi_l(\mu)$ will be,
\begin{equation}\label{phi for legendre1 without emission}
\begin{split}
\phi_l(\mu)
=& \mu^2-\frac{\mu}{2}\tilde{\omega_1}\eta_l(\mu)\int_0^1\frac{\eta_l(\mu')}{\mu+\mu'}\mu'^2d\mu'+\frac{\mu}{2}\frac{3\tilde{\omega_0}}{\zeta}\phi_l(\mu)\int_0^1\frac{\phi_l(\mu')}{\mu+\mu'}\mu'^2d\mu'\\
&+\frac{\mu}{2}\frac{3\tilde{\omega_2}}{4\zeta}\psi_l(\mu)\int_0^1\frac{\psi_l(\mu')}{\mu+\mu'}\mu'^2d\mu'
\end{split}
\end{equation}
\item The functional form of $\psi_l(\mu)$ will be,
\begin{equation}\label{psi for legendre1 without emission}
\begin{split}
\psi_l(\mu)
=& (\zeta-\mu^2) -\frac{\mu}{2}\tilde{\omega_1}\eta_l(\mu)\int_0^1\frac{\eta_l(\mu')}{\mu+\mu'}(\zeta-\mu'^2)d\mu'+\frac{\mu}{2}\frac{3\tilde{\omega_0}}{\zeta}\phi_l(\mu)\int_0^1\frac{\phi_l(\mu')}{\mu+\mu'}(\zeta-\mu'^2)d\mu'\\
&
+\frac{\mu}{2}\frac{3\tilde{\omega_2}}{4\zeta}\psi_l(\mu)\int_0^1\frac{\psi_l(\mu')}{\mu+\mu'}(\zeta-\mu'^2)d\mu'
\end{split}
\end{equation}
\item The intensity $I(0,\mu;\mu_0)$ will be,
\begin{equation}\label{I_(0) legendre without emission}
\begin{split}
&I(0,\mu;\mu_0)=\\ 
&\frac{\mu_0}{\mu+\mu_0}\frac{F}{4}[-\tilde{\omega_1}\eta_l(\mu)\eta_l(\mu_0) + \frac{3\tilde{\omega_0}}{\zeta}\phi_l(\mu)\phi_l(\mu_0) + \frac{3\tilde{\omega_2}}{4\zeta}\psi_l(\mu)\psi_l(\mu_0)\\
&+\{\tilde{\omega_1}(1+l\mu)(1+l\mu_0)-3\tilde{\omega_2}m^2\mu\mu_0\}H^{(1)}(\mu)H^{(1)}(\mu_0)\sqrt{(1-\mu^2)(1-\mu_0^2)}\cos(\phi-\phi_0)\\
&+\frac{3\tilde{\omega_2}}{4}H^{(2)}(\mu)H^{(2)}(\mu_0)(1-\mu^2)(1-\mu_0^2)\cos2(\phi-\phi_0)]
\end{split}
\end{equation}
The expression of $S_l^{(0)}(\mu,\mu')$ is same as given in eqn.(16) in \citet{horak1961diffuse} and $I(0,\mu,\mu_0)$ is equivalent with the eqn. (20) of \citet{madhusudhan2012analytic}

\end{enumerate}
\end{enumerate}

The equations from \eqref{only diffusion approximation}-\eqref{I_(0) legendre without emission} are the expected forms of those expressions derived in this paper when we neglect the \textit{atmospheric thermal emission} $B(T)$.

\section{Contribution of thermal emission:}\label{B(T) contribution}
The scattering functions calculated in this paper eqns.\eqref{scattering function isotropic1},\eqref{sx2} and \eqref{s rayleigh1} all contained the thermal emission in a general form as follows,
\begin{equation}\label{emsn efct on scat}
\frac{\mu\mu_0}{\mu+\mu_0}4U(T)f(\mu)
\end{equation}
Here, $f(\mu)$ is the distribution function depending on different phase functions given below,
\begin{equation}
\begin{split}
f(\mu)=&M(\mu)\hspace{0.5cm}for\hspace{0.5cm}p=\tilde{\omega_0}\\
=&\psi_a(\mu)\hspace{0.5cm}for\hspace{0.5cm}p=\tilde{\omega_0}(1+x\cos\theta)\\
=&\gamma(\mu)\hspace{0.5cm}for\hspace{0.5cm}p=\frac{3}{4}(1+\cos^2\theta)\hspace{1cm} and\\
&\hspace{2.25cm}  p = \tilde{\omega_0}+\tilde{\omega_1}P_1(\cos\Theta)+\tilde{\omega_2}P_2(\cos\Theta)
\end{split}
\end{equation}
The same similarity can be seen in  intensity $I(0,\mu;\mu_0)$ equations \eqref{I(0)_isotropic},\eqref{I0 for px} and \eqref{I_(0) Rayleigh} as,
\begin{equation}
\frac{\mu_0}{\mu+\mu_0} B(T)f(\mu)
\end{equation}
This $f(\mu)$ holds a general integral form,
\begin{equation}\label{f(mu) equation}
f(\mu) = 1+\frac{1}{2}\int_0^1 S^{(0)}(\mu,\mu')\frac{d\mu'}{\mu'}
\end{equation}
So the contribution of thermal emission to the intensity \textit{I} can be written explicitly as,
\begin{equation}\label{Contribution of Thermal emission}
B(T)f(\mu) = B(T) + \frac{1}{4\pi}\int_0^{2\pi}\int_0^1B(T)S(\mu,\phi;\mu',\phi')\frac{d\mu'}{\mu'}d\phi'
\end{equation}

In the same way, the contribution of thermal emission on scattering function $S$ can be expressed as,

\begin{equation}\label{Contribution of Thermal emission on S}
U(T)f(\mu) = U(T) + \frac{1}{4\pi}\int_0^{2\pi}\int_0^1U(T)S(\mu,\phi;\mu',\phi')\frac{d\mu'}{\mu'}d\phi'
\end{equation}
\begin{figure}[h!]\label{self scattering}
\begin{center}
\begin{tikzpicture}
\draw (0,0)--(9,0)node[right] {$\tau=0$};
\draw (0,-2)--(4.5,-2)node[above]{$(-\mu',\phi')$ direction};
\draw (4.5,-2)--(9,-2)node[right]{$\tau=\tau_1$};
\draw[<-] (4.5,0.9)node[above]{$B(T)$}--(4.5,0);
\draw[<-] (3.5,-1)--(4.5,0);
\draw[<-] (2.5,-1.5)--(3.5,-1);
\draw[<-] (3,0)--(2.5,-1.5);
\draw[->] (3,0)--(3,0.9);
\draw[<-] (6.5,-1)--(4.5,0);
\draw[<-] (7,0)--(6.5,-1);
\draw[->] (7,0)--(7,0.9);
\draw[->] (-0.1,0.2)--(-0.1,1)node[right]{$(+\mu,\phi)$ direction};
\end{tikzpicture}
\end{center}
\caption{This figure shows contribution of thermal emission along the $(+\mu,\phi)$ direction by direct as well as after scattering from $(-\mu',\phi')$ direction given by equation.\eqref{Contribution of Thermal emission}. Ray denoted as $B(T)$ will go directly along $(+\mu,\phi)$ direction from $\tau=0$ and other rays are initially along the directions $-\mu'$s and then scattered a number of times to finally come along $(+\mu,\phi)$ direction.}
\end{figure}

Equation \eqref{Contribution of Thermal emission} and \eqref{Contribution of Thermal emission on S} shows the total effect of thermal emission on diffuse reflection radiation and scattering function respectively. The first term on right hand side gives the planck emission along the direction $(+\mu,\phi)$. This is independent of direction. The second term gives the \textit{thermal emission reflected diffusely} along $(\mu,\phi)$ direction. Here the scattering function $S(\mu,\phi;\mu',\phi')$ scatters the thermal emission from $(-\mu',\phi')$ direction to $(\mu,\phi)$ direction (see fig. 2).

\section{Discussion:}\label{Discussion}

We derive the analytic solutions of diffuse reflection problem of semi-infinite homogeneous atmosphere introduced by \citet{Chandrasekhar} in presence of thermal emission using \textit{invariance principle method} \citet{ambartsumian1943cr}. In absence of thermal emission our treatment will reduce into only diffusely reflected case which has been previously studied by Chandrasekhar in a series of papers and tabulated in \cite{Chandrasekhar},\citet{horak1950diffuse},\citet{chandrasekhar1947radiative1}. We mention that those tabulated expressions are exactly same with the reduced form of our generalized equations as shown here in \eqref{only diffusion approximation}-\eqref{I_(0) Rayleigh without emission}.

The reduction of the scattering integral equations have been done for three different phase functions other than the isotropic case. In each case the azimuth independent component of scattering function $S^{(0)}$ is affected by the thermal emission and get modified from that of Chandrasekhar. Other $S^{(i)}$ and $H^{(i)}$ terms (with $i\neq0$) for Asymmetric scattering, Rayleigh scattering and the scattering for the phase function $\tilde{\omega_0}+\tilde{\omega_1}P_1(\cos\Theta)+\tilde{\omega_2}P_2(\cos\Theta)$ remains unaffected. They are exactly the same as given in \citet{Chandrasekhar},\citet{horak1961diffuse}.This is because we consider the atmospheric emission is thermal emission only. Now, as the thermal emission $B(T_\tau)$ is isotropic in nature (see fig. 1), so the emission contribution will always affect the isotropic (i.e. azimuth independent) terms of the scattering function and all other constituent terms remains unaffected as expected. 

This comparison shows that, the emission alongwith the diffusion reflection is more general than the case of only diffusion reflection considered in \citet{Chandrasekhar}. We considered here three special cases (other than isotropic scattering) of scattering phase function $p(\mu,\phi;\mu',\phi')$ by expanding it in terms spherical harmonics. We can expand it even in more general form of Legendre polynomial introduced in \cite{Chandrasekhar} as,
\begin{equation}
p(\cos \Theta) = \sum_{l=0}^\infty \tilde{\omega_l}P_l(\cos\Theta)
\end{equation}
where $\tilde{\omega_l}$'s are constants. Then the corresponding \textit{Scattering Function} will be represented as follows,
\begin{equation}
S = \sum_{l=0}^\infty \tilde{\omega_l}S^{(l)}P_l(\cos \theta)
\end{equation}

From our study we can directly imply that, the effect of thermal emission $B(T_\tau)$ will contribute in $S^{(0)}$ terms only and all other terms remains unaffected for $l\neq0$ of whatever expansion is considered. To see the emission effect in higher degree of scattering functions (i.e. $S^{(1)}, S^{(2)}$... etc) one can consider the asymmetry in atmospheric emission, which is different from planck emission.

The intensity $I(0,\mu;\mu_0)$ derived in this paper always includes a thermal emission of whatever scattering phase function is considered as shown in section~\ref{B(T) contribution}. Thus the blackbody emission $B(T)$ always adds some radiation to $I(0,\mu,\mu_0)$, multiplied by the function $f(\mu)$. The explicit form of $f(\mu)$ reveals that the contribution of thermal emission contains direct emission along $(\mu,\phi)$ as well as diffusely scattered radiation from other directions to $(\mu,\phi)$ (see fig. 2).

The scattering function $S^{(0)}(\mu,\mu_0)$ affected by the thermal emission in terms of $U(T)$ as shown in eqn. \eqref{emsn efct on scat}. As, the multiplication factor $U(T)$ is the ratio of planck emission and incident flux $\pi F$, so it can be stated that \textit{the effect of thermal emission on scattering function is inversely  proportional to the incident flux $\pi F$ and directly proportional to the blackbody emission from the corresponding layer.} Thus, when the irradiation flux $\pi F$ increases but the planck emission remains fixed then the relative effect of thermal emission on scattering function is supressed.

Chandrasekhar's semi-infinite atmosphere model is used for a large number of cases. For example \citet{king1963greenhouse} solved the green house effect of semi-infinite atmosphere, \citet{dubus1986theoretical} used the model to evaluate ion induced secondary electron emission whereas \citet{madhusudhan2012analytic} analytically model exoplanetary albedo, phase curve and polarization of reflected light using the direct results derived in \citet{Chandrasekhar}. We have shown that the specific intensity and scattering functions are underestimated while not considering the thermal emission. Thus to get accurate estimations the inclusion of thermal emission is important while using the diffuse reflection model of semi-infinite atmosphere.

The inclusion of atmospheric emission in terms of planck function to semi-infinite atmosphere problem is the first step towards the generalization of Chandrasekhar's treatment. Though there are some limitations to this model. For instance, in case of exoplanetary atmosphere the assumption of Local thermodynamic equilibrium is not valid at upper atmospheric region\citet{seager2010exoplanet}. Thus, the atmospheric emission will show a departure from pure blackbody emission and should be modified by other emission effects. One can treat this problem for atmospheric re-emission case \citet{chakrabarty2020effects}, by replacing $\beta(\tau,\mu,\phi)=(1-\tilde{\omega_0})B(T_\tau)$ in eqn.\eqref{source function for emission+scattering1} and all the results follows accordingly.

Again, we considered the low scatteing limit ($\kappa>>\sigma$) in this work for which $\beta$ entirely boils down into the planck function B(T). To remove this restriction, the thermal emission B(T) can be replaced by, $\boldsymbol{\frac{\kappa}{\chi}}$B(T) which modifies the results.

 Also we assume that the atmospheric emission is planck emission only, which is isotropic in nature. This is oversimplification of the practical problem. The anisotropic effect of atmospheric emission can be included by taking the fourier expansion,
$$\beta(\tau,\Omega)=\sum_{m=0}^n \beta_m(\tau,\mu;\mu_0)\cos(\phi-\phi_0)$$
as given in \citet{bellman1967chandrasekhar}. In that case not only $S^{(0)}$ but all $S^{(i)}$ terms of the scattering function as well as the $I(0,\mu,\mu_0)$ will be modified. This can be a more practical approach to the problem and much rigorous calculations are needed.

The phase functions considered here, all have analytical forms given in \citet{Chandrasekhar}. But for more realistic problems of single or direct scattering with a forward scattering effect, \citet{henyey1941diffuse} introduced a phase function, 

\begin{equation*}
p(\cos\Theta) = \frac{1-g^2}{(1+g^2-2g\cos\Theta)^\frac{2}{3}}
\end{equation*}
where, $g\in[-1,1]$ is the asymmetry parameter. This type of phase function has been well studied for reflected spectroscopy using numerical analysis \citet{batalha2019exoplanet}. To treat this in our semi-infinite atmosphere diffuse reflection problem one should shift from analytical treatment to numerical one.

Finally, we did not include the polarization effect to our calculations. Following \citet{Chandrasekhar}, we can say that all of our results will be valid while including polarization effect with some replacements as follows. The intensity \textit{I} will become a vector \textbf{I} whose components are the stokes parameters. The phase function \textit{p} and scattering function \textit{S} will be replaced by the analogous phase matrix \textbf{P} and scattering matrix \textbf{S}. In that scenario, polarization effect in semi-infinite atmosphere problem with atmospheric emission can be studied.

\textit{Acknowledgement:} SS is thankful to M. Singla for her valuable discussion and suggestions. SS also acknowledges the support by DST for funding this project.
\bibliography{paper1}
\bibliographystyle{aasjournal}
\end{document}